\documentstyle[epsf]{article}
\begin{document}
\twocolumn
\noindent {\bf Developments in High Energy Neutrino 
Astronomy\footnote{To appear in {\it Europhysics News}, 1999}  
}\\
R.J. Protheroe, University of Adelaide, Australia \\

Nature produces cosmic ray particles, probably protons, with
energies well above $10^{20}$ eV -- how are they produced?  Where
do they come from?  Gamma rays with energies above $10^{13}$ eV
are produced in jets of active galaxies -- are these produced by
energetic electrons or protons?  What is the correct model of
Gamma Ray Bursts?  These are just some of the fundamental
questions in high energy astrophysics to be answered by
observations made with large area neutrino telescopes.

When gamma rays result from hadronic interactions of protons
neutrinos are also produced, but they are not produced when
energetic electrons Compton scatter X-rays to gamma-ray energies.
So neutrino observations may distinguish between models of active
galactic nuclei.  Similarly, models for the origin of the highest
energy cosmic rays (almost certainly extragalactic) --
acceleration of protons in hot spots of giant radio galaxies,
acceleration by Gamma Ray Bursts, decay of massive X particles
produced by topological defects -- may be distinguished as they
have very different neutrino signatures, as do different models
for Gamma Ray Bursts.

One big advantage of neutrino astronomy is that because of their
low interaction cross section neutrinos can escape from regions
opaque to photons, but this is also their biggest disadvantage as
most neutrinos pass unobserved through the Earth.

Predicted fluxes and detection rates are very low, and telescopes
of area approaching 1 km$^2$ are needed to detect diffuse
background intensities.  More than three decades ago, Russian and
American physicists thought of instrumenting a huge volume of
water with photomultiplers to look for Cherenkov light produced
by an upward-going muon produced by an interaction below the
detector of a muon-neutrino which had passed upwards through the
Earth.

The AMANDA telescope located at the South Pole has recently
detected sixteen upward-going neutrino events with energies in
the range $10^{11}$ eV to $10^{12}$ eV.  The detectors used
consisted of ten ``strings'' of photomultipliers (a total of 289)
placed in deep vertical holes in the transparent Antarctic ice.
Each string extends from 1.5 km to 2 km depth, and the holes are
spread over $10^4$ m$^2$ at the surface.  All sixteen events are
probably produced by cosmic rays interacting with air nuclei in
the Northern Hemisphere, but they are detected well above any
background.  The neutrino-induced muon tracks are clearly seen
and the angular resolution is very good, currently about
2$^\circ$.  One of the events has a muon track which is
nearly vertical and is almost coincident with String 6 -- shown in
the diagram (provided by Francis Halzen of the University of
Wisconsin) where the signal in each detector is indicated by the
size of the circle.  The graph shows the time each
photomultiplier triggered plotted against its depth, giving a
perfect correlation for detectors on String 6, and showing the
muon's speed was close to the speed of light.  This exciting
result shows that the goal of constructing viable telescopes for
high energy neutrino astronomy is achievable at reasonable cost.

\begin{figure}
\epsfysize12.3cm\epsfbox{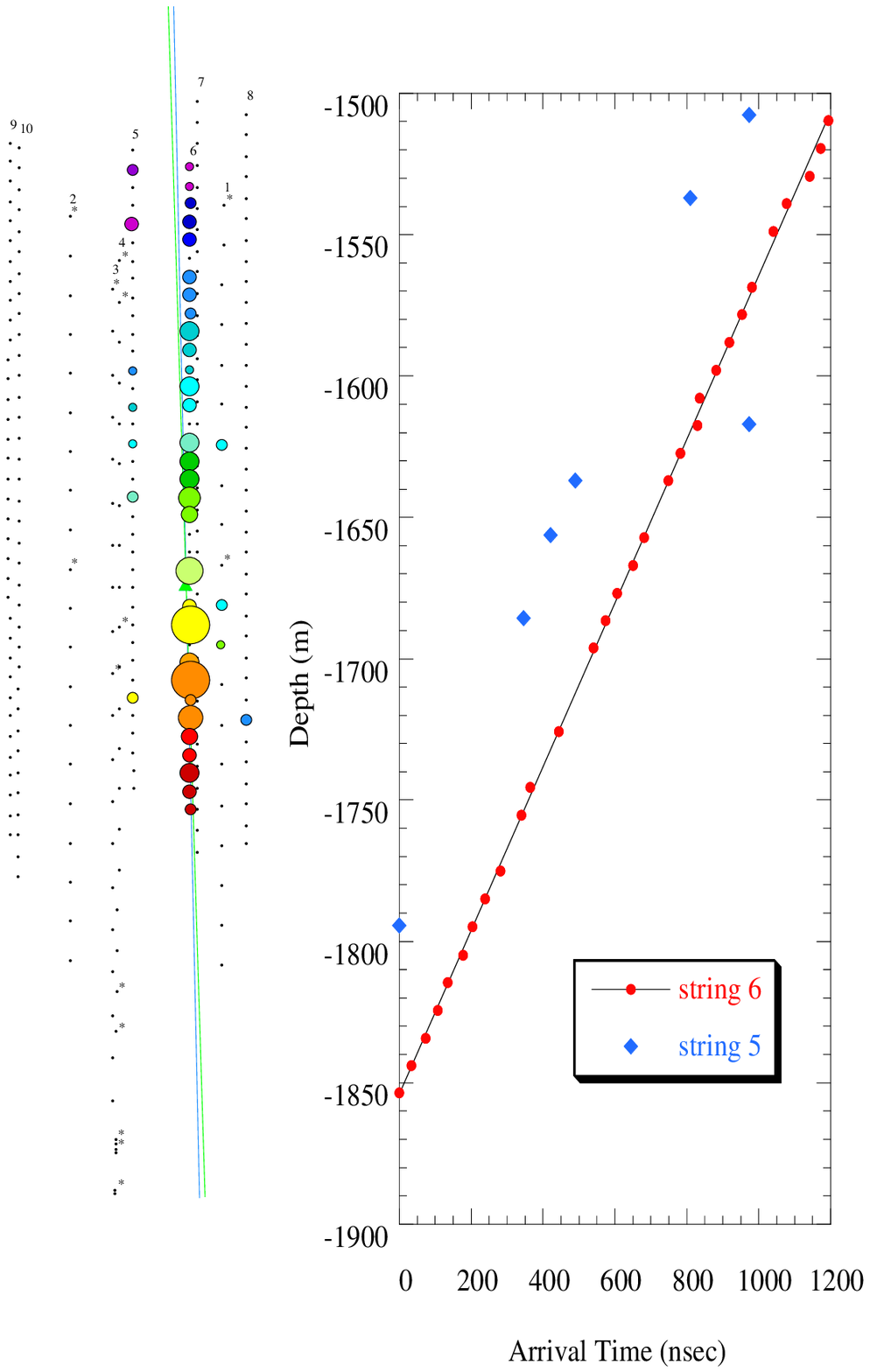}
\end{figure}

\end{document}